\documentclass[twocolumn, prb,preprintnumbers,amsmath,amssymb,floatfix]{revtex4}
\usepackage[linktocpage,bookmarksopen,bookmarksnumbered]{hyperref}

\usepackage{dcolumn}

\usepackage{amsmath,graphics,epsfig,color,verbatim,ulem}
\usepackage{braket}

\renewcommand{\vr}{{\mathbf{r}}}

\usepackage{tabularx,ragged2e,booktabs,caption}
\newcolumntype{C}[1]{>{\Centering}m{#1}}

\usepackage{multibib}

\begin{document}

\setlength{\pdfpageheight}{\paperheight}
\setlength{\pdfpagewidth}{\paperwidth}

\title{Mott Transition and Magnetism in Rare Earth Nickelates and its
  Fingerprint on the X-ray Scattering}
\author{Kristjan Haule*}
\author{Gheorghe L. Pascut}
\affiliation{Department of Physics and Astronomy, Rutgers University, Piscataway, NJ 08854, United States.}
\date{\today}
\begin{abstract}
  The metal-insulator transition (MIT) remains among the most
  thoroughly studied phenomena in solid state physics, but the
  complexity of the phenomena, which usually involves cooperation of
  many degrees of freedom including orbitals, fluctuating local
  moments, magnetism, and the crystal structure, have resisted
  predictive \textit{ab-initio} treatment.  Here we develop
  \textit{ab-initio} theoretical method for correlated electron
  materials, based on Dynamical Mean Field Theory, which can predict
  the change of the crystal structure across the MIT at finite
  temperature. This allows us to study the coupling between
  electronic, magnetic and orbital degrees of freedom with the crystal
  structure across the MIT in rare-earth nickelates.  We predict the
  electronic free energy profile of the competing states, and the theoretical
  magnetic ground state configuration, which is in agreement with
  neutron scattering data, but is different from the magnetic models
  proposed before.  The resonant elastic X-ray response at the K-edge,
  which was argued to be a probe of the charge order, is
  theoretically modelled within the Dynamical Mean Field Theory,
  including the core-hole interaction.  We show that the line-shape of
  the measured resonant elastic X-ray response can be explained with
  the "site-selective'' Mott scenario without real charge order on Ni
  sites.
\end{abstract}
\maketitle




Metal-insulator transition (MIT) in transition metal oxides is usually
associated with a large Hubbard Coulomb interaction $U$ on transition
metal ion, which strongly impedes electron motion, as it costs an
energy $U$ to add an extra electron to any given site. Consequently
electrons become localized on the transition metal ion, and hence form
a fluctuating moment, which possesses a large entropy that is being
released at low temperature by emergence of a long range magnetic
order. But most MITs are much more complex than that, and require
cooperation of several degrees of freedom, including the subtle change
of the crystal structure to tune the hybridization with the oxygen,
the modulation of the strength of the fluctuating moments and orbital occupations. In
\textit{ab-initio} modeling, this requires one to optimize the crystal
structure to the correlated electronic state as an external parameter is varied.

\begin{figure*}[bht]
\includegraphics[width=0.8\linewidth]{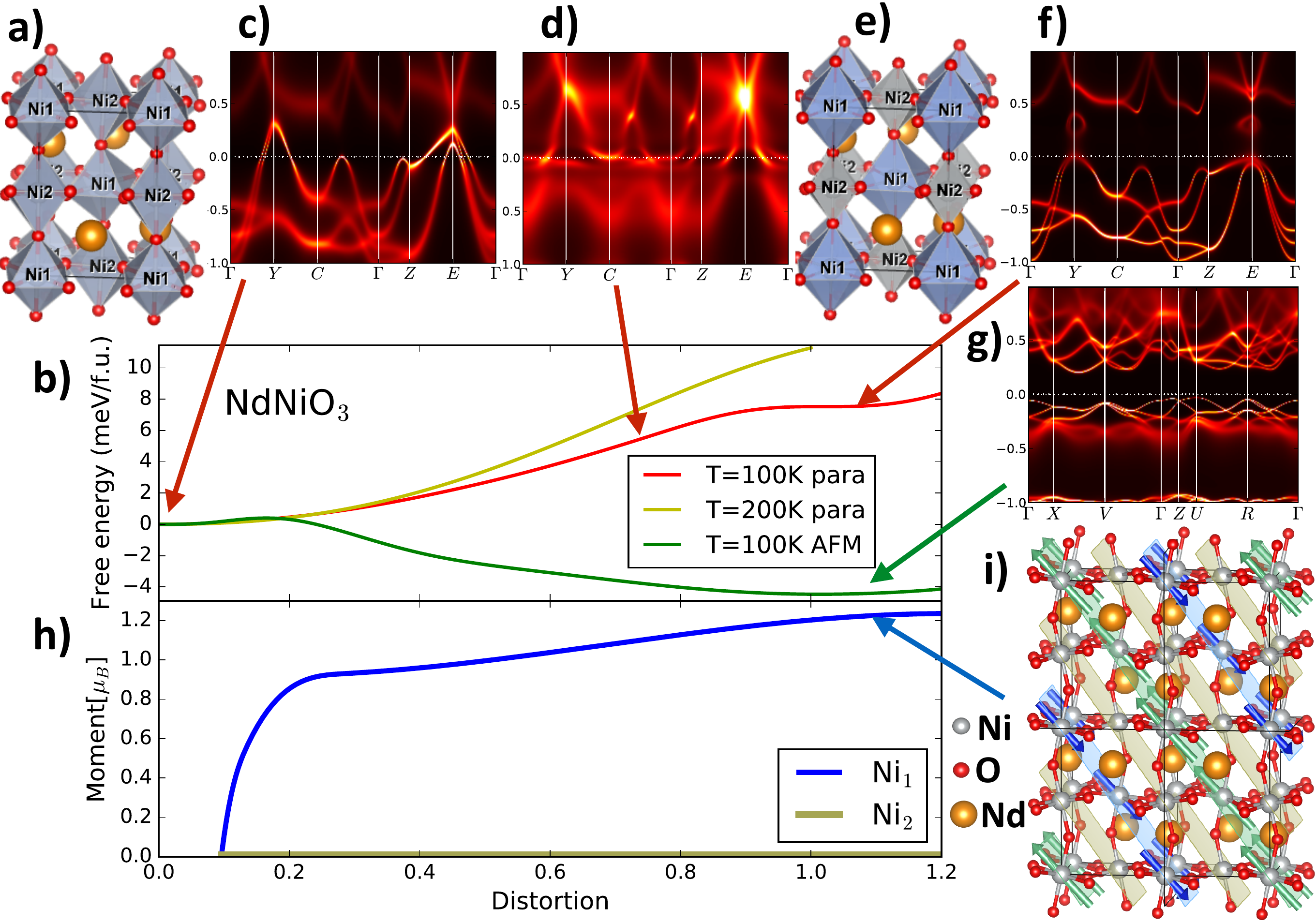}
\caption{ \textbf{Energetics and magnetism of NdNiO$_3$:}  
a) The crystal structure of the metallic NdNiO$_3$ stable above $T\gtrsim 200\,$K.
b) The electronic free energy of theoretical paramagnetic solution, and
antiferromagnetic (AFM) solution as a function of distortion (i.e., linear
interpolating between two local minima). 
c) spectral function of the paramagnetic metallic state stable at high
$T$, d) spectral function of a metastable state at 80\% of distortion,
e) distorted structure of the insulating state, f) spectral function
of the paramagnetic insulating solution at the low $T$ equilibrium structure,
g) spectral function of the AFM solution h) magnetic
moment of the two nickel ions in the AFM state, i) theoretically
determined magnetic configuration of the ground state. The planes in
(1,0,1) direction contain three types of Ni ions: the green (blue) planes
contain Ni$_1$ atoms with magnetic moments pointing up (down), while the yellow planes contain
Ni$_2$ atoms which carry no magnetic moment.
}
\label{fig1}
\end{figure*}

The MIT in $R$NiO3~\cite{firstRNiO3} is accompanied by the structural
transition in which the high-temperature metallic phase, with the
orthorhombic (Pbnm) structure (see Fig.~\ref{fig1}a), is transformed
to the low-temperature insulating phase of monoclinic (P2$_1$/n)
structure. In the latter, the alternating NiO$_6$ octahedra
are expanded and compressed in a rocksalt-pattern distortion
(see Fig.~\ref{fig1}e).~\cite{Ni_HT_struct,YNiO3_structure} The transition
is accompanied by the antiferromagnetic ordering, which occurs
simultaneously with the MIT in Nd and Pr compound ($R$=Nd,Pr) and at lower
temperature for the smaller rare-earth ions ($R$ = Sm and beyond).
%
%



%
The structurally distorted monoclinic ground state is very susceptible to
small changes of external parameters and can be tuned by
pressure~\cite{firstRNiO3},
strain~\cite{strain1,strain2,strain3,strain4}, reduced
dimensionality~\cite{dim1,dim2} or by layering it in
heterostructures~\cite{heteros1,heteros2,heteros3}, hence it has
attracted a lot of attention recently.
%

The leading interpretation for the origin of the MIT is a charge
disproportionation (CD) on the Ni sites, in which Ni$^{3+}$ ions
disproportionate into sites with excessive and deficient charge
($3d^7 3d^7\rightarrow 3d^{7+\delta} 3d^{7-\delta}$). Such charge
order would result in different energy positions of core levels on the
two inequivalent Ni
ions due to electrostatic effect, which can be probed by the
hard resonant elastic X-ray scattering (RXS) through measuring the $1s$ to $4p$
transitions. In Ref.~\onlinecite{Xray1} it was estimated that 
the charge order is approximately $2\delta\approx
0.42\,e$.~\cite{Xray1}, based on the $1s-4p$
energy difference of around 0.9$\,eV$ for the two inequivalent Ni
ions.
Similar conclusion was
reached by numerous other resonant scattering
techniques.~\cite{YNiO3_structure,charge_order2,charge_order3,charge_order4,charge_order5}.
This view has been challenged theoretically, since the \textit{ab-initio}
calculations predict very small rearrangement of electronic charge
across the transition.~\cite{Picket,Pascut}
On the other hand, the weak coupling theories are supportive of this picture, but also
emphasize the cooperation of charge and spin-density wave, with the
latter being the driving force of the MIT in NdNiO$_3$ and
PrNiO$_3$.~\cite{SDW1,SDW2}

The alternative explanation posits that Ni experience a "negative
charge transfer energy" and consequently is found in a very different
$d^8$ valence state with compensating holes on the oxygen
sites. ~\cite{negative_charge_transfer,referee} The compressed octahedra
contains Ni $d^8$ ion and two ligand holes and the three bind into net
zero spin producing unusual continuum of particle-hole
excitations~\cite{referee}, while in the expanded octahedra Ni $d^8$
ion is in
high-spin $S=1$ state.~\cite{Sawatzky_recent} Theoretical studies
which assume such negative charge transfer
energy
found a novel state dubbed "site-selective'' Mott phase.~\cite{Hyowon,Hyowon2}  In this picture the
Ni ions in the expanded octahedra undergo the usual Mott transition
with two holes on Ni giving rise to a very strong $S=1$ local moment,
while the electrons in the compressed octahedra bind with the
(primarily oxygen) states near the Fermi level, and the resulting
bonding-antibonding gap opens up, similarly to the band gap of a Kondo
insulator.~\cite{Hyowon,Antoine}

Although this very appealing picture is accumulating strong support,
many fundamental questions remain: i) How to reconcile the RXS 
experiments, which require CD,
with the picture of "site-selective'' Mott transition.  
ii) In the seminal work on "site-selective'' Mott
transition~\cite{Hyowon}, the physical $d^8$ valence was reached by
adjusting the onsite energy  through
an ad-hoc double-counting adjustment, in which Coulomb $U$ in the
interaction and double-counting were different, in order to reach the
"negative charge transfer energy" regime. Similarly, in cluster
calculations~\cite{Sawatzky_recent} the model parameters are chosen
such that Ni is found in $3d^8$ configuration. Since the exact
double-counting between the Dynamical Mean Field Theory and Density
Functional Theory has been derived recently~\cite{exactDC}, the
assumption of nickel $3d^8$ valence can now be checked without
resorting to any \textit{a-priori} assumption on Ni valence.  iii) The
propagating vector of the antiferromagnetic order has been
unambiguously determined by the neutron
scattering~\cite{first_neutron1}, while the precise magnetic
configurations was challenging to constrain, and different experiments
were interpreted in terms of conflicting models of
collinear~\cite{first_neutron2,morder_late} and
non-collinear~\cite{morder_early} magnetic order. On the other hand,
the \textit{ab-initio} electronic structure methods are not supportive of so
far proposed models, and suggest that ferromagnetic state is favored
compared to proposed antiferromagnetic
orders~\cite{Hyowon,Hyowon2}. iv) Many experiments on the Pr and Nd
compound~\cite{Keimer_RXS} were interpreted in terms of an itinerant
picture~\cite{SDW1,SDW2} in which the spin-density wave drives the
MIT. An important question arises: is the magnetic long range order
necessary for the MIT in these systems, or, is the Neel order just a
consequence of the MIT and it is just a way in which the existing
local moments release their entropy.
 
To address these issues, we use \textit{ab-initio} theoretical method
for correlated electron materials, based on combination of dynamical
mean field theory (DMFT) and density functional theory
(DFT)~\cite{Kotliar_rmp06}, in its real space embedded form~\cite{Haule_prb10}, which
avoids downfolding.~\cite{DMFTcode} To address the issue of Ni valence, we use
recently derived exact double-counting between the DFT and DMFT
methods~\cite{exactDC}. To successfully address the energetics of
different competing states and to determine the ground state of the
system, it is crucial to theoretically determine the optimized crystal
structure, and for this we use recent implementation of forces within
DFT-DMFT.~\cite{forces}

We checked that within this theoretical approach LaNiO$_3$
remains paramagnetic metal at least down to 50$\,K$ and does not show
any sign of long range order, in agreement with experiment. On the
other hand NdNiO$_3$ shows the existence of three phases, the
paramagnetic insulating, the antiferromagnetic insulating and the
paramagnetic metallic phase. In Fig.~\ref{fig1}
we show the energetics of these phases as predicted by
the theory. 
The paramagnetic metallic phase is stable above
200$\,K$. Its spectra is plotted in Fig.~\ref{fig1}c. The
crystal structure in this phase is fully relaxed within the DFT-DMFT
theory, and its predicted structural parameters are in excellent
agreement with the experiment (see table~\ref{tab:struct}). For
comparison we show the GGA relaxation of the structure, which shows
three times larger disagreement with experiment. 
When the temperature
is lowered to above 100$\,$K a first sign of structural instability
occurs, as shown in Fig.~\ref{fig1}b. The electronic free energy curve of the
paramagnetic phase develops a local minimum in the P2$_1$/n structure,
where oxygen octahedra around Ni$_1$ sites are expanded, and octahedra around
Ni$_2$ sites are compressed. Using the technology to calculate forces~\cite{forces}, we
optimized the structural parameters in this phase (see chapter I in the
supplementary information~\cite{suppl}).
In the local minimum, the Mott gap opens up on
Ni$_1$ atom, while Ni$_2$, through strong hybridization with the
environment, splits bands such that the band gap opens at the Fermi
level, all consistent with the "site-selective'' Mott
transition scenario~\cite{Hyowon} (see Fig.~\ref{fig1}f). Just slightly away
from this local minimum (80-90\% distortion), the insulator breaks
down and strongly incoherent metallic state appears
(Fig.~\ref{fig1}d). 

{\scriptsize
\centering
\captionof{table}{
\footnotesize
Optimized atomic positions in the metallic and
  insulating state of NdNiO$_3$. Experimental structure is from
Ref.~\onlinecite{Ni_HT_struct}. The GGA and GGA+U structure is from
Ref.~\onlinecite{Keimer_RXS}.}
\label{tab:struct} 
\begin{tabular}{l|c|c|c|c}
Pbnm          & Exp. & DMFT-PARA  & GGA \\
\hline
Ni       & (0.000, 0.000, 0.500) & (0.000, 0.000, 0.500) & (0.000, 0.000, 0.500)\\
O$_1$ & (0.216, 0.287, 0.539) & (0.214, 0.287, 0.539) & (0.207, 0.294, 0.547)\\
O$_2$ & (0.569, 0.490, 0.750) & (0.573, 0.490, 0.750) & (0.591, 0.477, 0.750)\\ 
Nd      & (0.496, 0.035, 0.750) & (0.491, 0.044, 0.750) & (0.488, 0.058, 0.750)\\
\multicolumn{2}{l|}{$\sqrt{\braket{(\vr-\vr_{exp})^2}}$} & 0.0056     &  0.0190\\
\hline
P2$_1$/n         & Exp &  DMFT-AFM& GGA+U \\
\hline
Ni$_1$&(0.000, 0.000, 0.000) &(0.000, 0.000, 0.000)&(0.000, 0.000, 0.000)\\
Ni$_2$&(0.000, 0.000, 0.500) &(0.000, 0.000, 0.500)&(0.000, 0.000, 0.500)\\
O$_1$ &(0.575, 0.487, 0.752) &(0.574, 0.489, 0.750)&(0.595, 0.475, 0.755)\\ 
O$_2$ &(0.214, 0.276, 0.527) &(0.209, 0.285, 0.540)&(0.198, 0.291, 0.549)\\ 
O$_3$ &(0.719, 0.204, 0.447) &(0.717, 0.210, 0.460)&(0.711, 0.198, 0.452)\\
Nd      &(0.493, 0.039, 0.750) &(0.493, 0.044, 0.750)&(0.489, 0.056, 0.750)\\
\multicolumn{2}{l|}{$\sqrt{\braket{(\vr-\vr_{exp})^2}}$} &0.0091&     0.0180\\
\end{tabular}
\par
\bigskip
}
In the Pbnm structure (zero distortion in Fig.~\ref{fig1}b,h) the
fluctuating moments are present, but they are not strong enough to
allow for the long range magnetic order, hence the system resolves its
excess entropy in the Fermi liquid state at low temperature.  Once the
Ni$_1$ hybridization is reduced a bit due to small increase of the
oxygen octahedra (around 10\% distortion), the correlations on Ni$_1$
become strong enough so that the static magnetic moment appears (see
Fig.~\ref{fig1}h). These correlations are primarily driven by the
strong Hund's coupling on Ni ion, which aligns two holes on the Ni$_1$
site, but the static ordered moment is only about 2/3 of the maximum
moment for spin $S=1$ state.  The resulting magnetic configuration,
predicted by the present theory, is displayed in Fig.~\ref{fig1}i. The
magnetic unit cell quadruples, and the magnetic moment of Ni$_1$ ions
in the parallel planes in (1,0,1) direction are ferromagnetically
aligned.  The static moments on Ni$_2$ however remains exactly zero,
as the fluctuating moment on Ni$_2$ gets even reduced in the distorted
(P2$_1$/n) structure, and the 
strong bonding with the surrounding oxygen concomitant with the
appearance of the band gap, prevents any
static moment on that site.  Every second Ni plane thus carries
magnetic moment, and those Ni$_1$ planes couple antiferromagnetically.
This ordering of moments on Ni$_1$ sublattice
coincides with the proposed model deduced from the neutron
scattering~\cite{first_neutron2} and resonant soft X-ray
diffraction~\cite{morder_early}, but it differs from both models due
to Ni$_2$ sites. In the proposed neutron-scattering model~\cite{first_neutron2}
Ni$_2$ moments were arranged antiferromagnetically within a single
$(1,0,1)$ plane, while in soft X-ray diffraction
model~\cite{morder_early}, Ni$_2$ moments were arranged
ferromagnetically, but 90 degrees rotated with respect to Ni$_1$
moments, so that the resulting magnetic structure is
non-collinear. The magnetic long-range solutions in the DFT-DMFT
theory can not sustain finite static moment on Ni$_2$, and we show in
chapter V of the supplementary~\cite{suppl} that the theoretical magnetic
configuration fits the neutron scattering data as good as the proposed
model of Ref.~\onlinecite{first_neutron2}. Our proposed magnetic
configuration is also consistent with the inelastic neutron scattering
result, which showed that all Nd ions experience similar Weiss field.~\cite{inelastic_neutron}

Finally, the gain in free energy is considerable once the magnetic
long range order is turned on, hence this magnetic order displayed in
Fig.~\ref{fig1}i is the theoretical ground state of the displayed unit
cell. Table~\ref{tab:struct} lists the optimized structure in the
magnetic state, which shows almost no difference as compared to the
paramagnetic structure in P2$_1$/n symmetry (given in chapter III of the supplementary
material~\cite{suppl}).  From Fig.~\ref{fig1}b we can also conclude that the
magnetism is not necessary for the metal-insulator transition, but in
NdNiO$_3$, this paramagnetic insulator appears metastable, and energy
gain due to long range magnetic order helps to stabilizes the
insulating state. In supplementary material (chapter II) we show that for smaller
rare earth ion (LuNiO$_3$) the paramagnetic insulating state is stable
at 100$\,K$ in the absence of magnetism. Magnetism is thus just an
efficient way to release the large entropy of fluctuating moment on
Ni$_1$ sites, which are formed with a help of much stronger Hund's
coupling mechanism.

\begin{figure}[bht]
\includegraphics[width=0.99\linewidth]{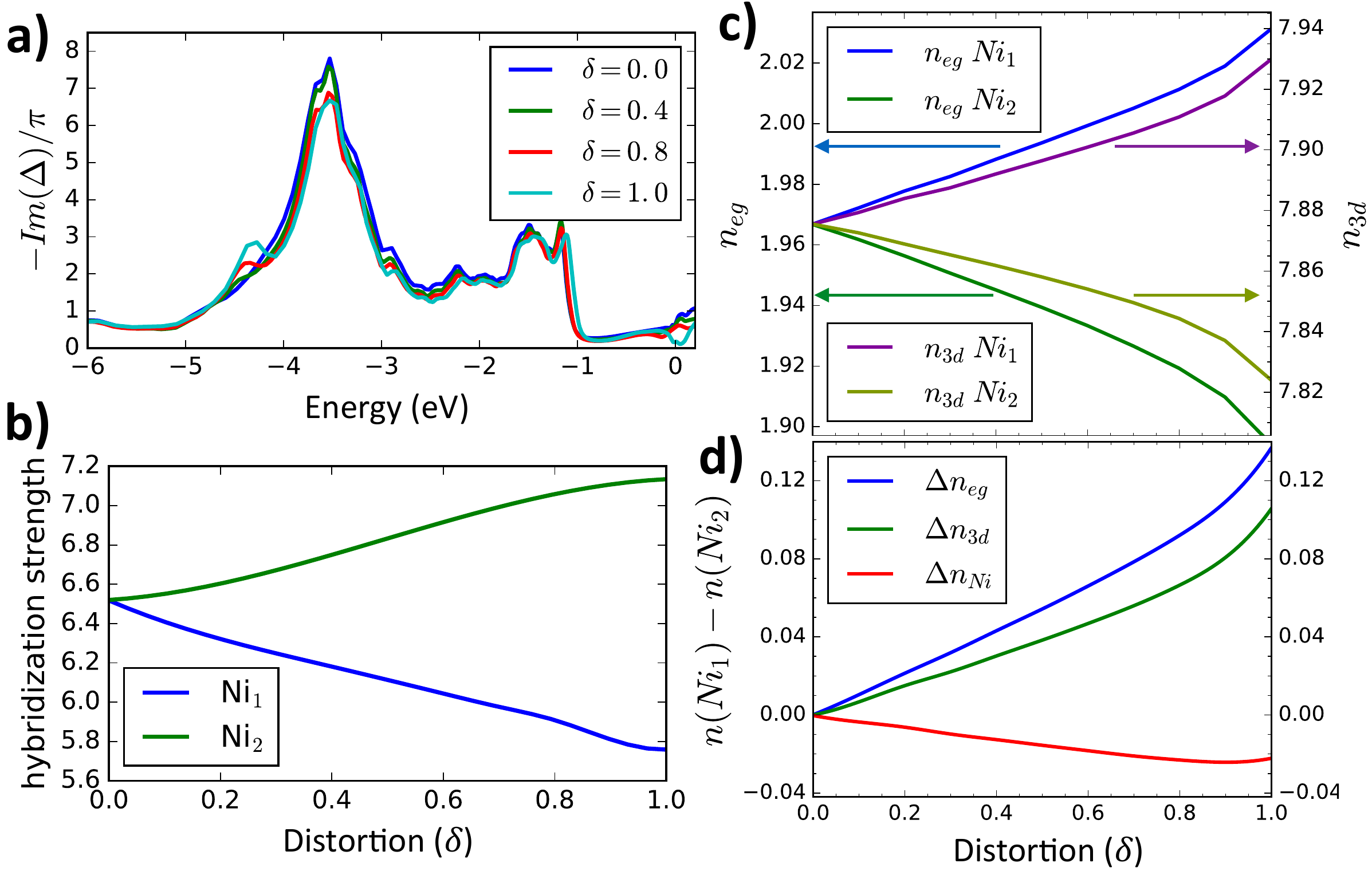}
\caption{ \textbf{Hybridization and charge of Ni ions:}  
a) Energy dependent hybridization function of the Ni$_1$ ion at few
values of the distortion parameter $\delta\in (0,1)$. b) The integral
of the hybridization function (in the displayed energy window) as a function
of distortion parameter $\delta$. c) The charge density on the two Ni
atoms corresponding to the $e_g$ orbitals and the entire $3d$
shell. d) The difference of the charge between Ni$_1$ and Ni$_2$ in
the $e_g$ orbital, in the $3d$ shell, and in the entire muffin-tin
sphere corresponding to Ni atoms.
}
\label{fig2}
\end{figure}
While the large Hund's coupling is essential for the appearance of
strong local moments on Ni$_1$ sites, the MIT in these materials is tuned
by the reduced hybridization on Ni$_1$ sites, displayed in
Fig.~\ref{fig2}a and b. It decreases for about 10\% in the
bond-disproportionate structure, and this is sufficient for a Mott
localization of electrons on Ni$_1$ site. Notice that the largest
contribution to the hybridization comes from nickel-oxygen overlap, and its
reduction is mostly concentrated at the energy of the center of
the oxygen states (see Fig.~\ref{fig2}a around -3.5$\,$eV).
On Ni$_2$ sites however, the
hybridization increases almost as much as it decreases on Ni$_1$
sites (see Fig.~\ref{fig2}b), but because the hybridized Ni$_2$ and oxygen states in this
crystal structure form a band insulator, this increased overlap does
not collapse the insulating gap. On the basis of this calculation, we
predict that the material would become a canonical Mott insulator if
hybridization on both sites gets as small as on Ni$_1$ in P2$_1$/n
structure, which might be possible to achieve in some thin
heterostructures of this material.~\cite{heteros3} 

In Fig.~\ref{fig2}c we display the electron charge on Ni ions versus
distortion, as obtained by projecting the electron charge to a
muffin-thin sphere around each Ni atom of size $\approx 2\,$a.u.  We
notice that there are approximately $2\,$eg electrons on each site,
and approximately $8\,$ electrons in the $3d$ shell, which corresponds
to Ni $3d^8$ configuration, as previously postulated in
Refs.~\onlinecite{Hyowon,Antoine,Zawatsky_nice}, and hence our exact
double-counting thus proves the correctness of the negative charge
transfer picture for these nickel compounds.
We also notice that Ni$_1$
(Ni$_2$) sites with large (small) octahedra gain (loose) some $eg$
electronic charge with structural distortion, and the difference of
the $eg$ charge on the two Ni sites becomes of the order of $0.14\,$electrons
in equilibrium P2$_1$/n structure. However, the electronic charge in the
entire $3d$ shell differs only for $\approx 0.1$electrons, and when all charge
inside muffin-thin sphere on Ni is counted, the charge difference is
negligible. Moreover, if we were to construct a very low energy model
comprised of only the lowest energy bands, we would completely
eliminate all Ni$_1$ states, as they are pushed to high energy Hubbard
bands, and we would conclude that all low energy holes come from
Ni$_2$ sites.  Hence, we can conclude that the appearance of charge
order depends on the type of model considered, and while there is no
real charge difference in the spheres around each Ni, the low energy
models comprised on Ni eg-states only, should allow for charge order.

\begin{figure}[bht]
\includegraphics[width=0.99\linewidth]{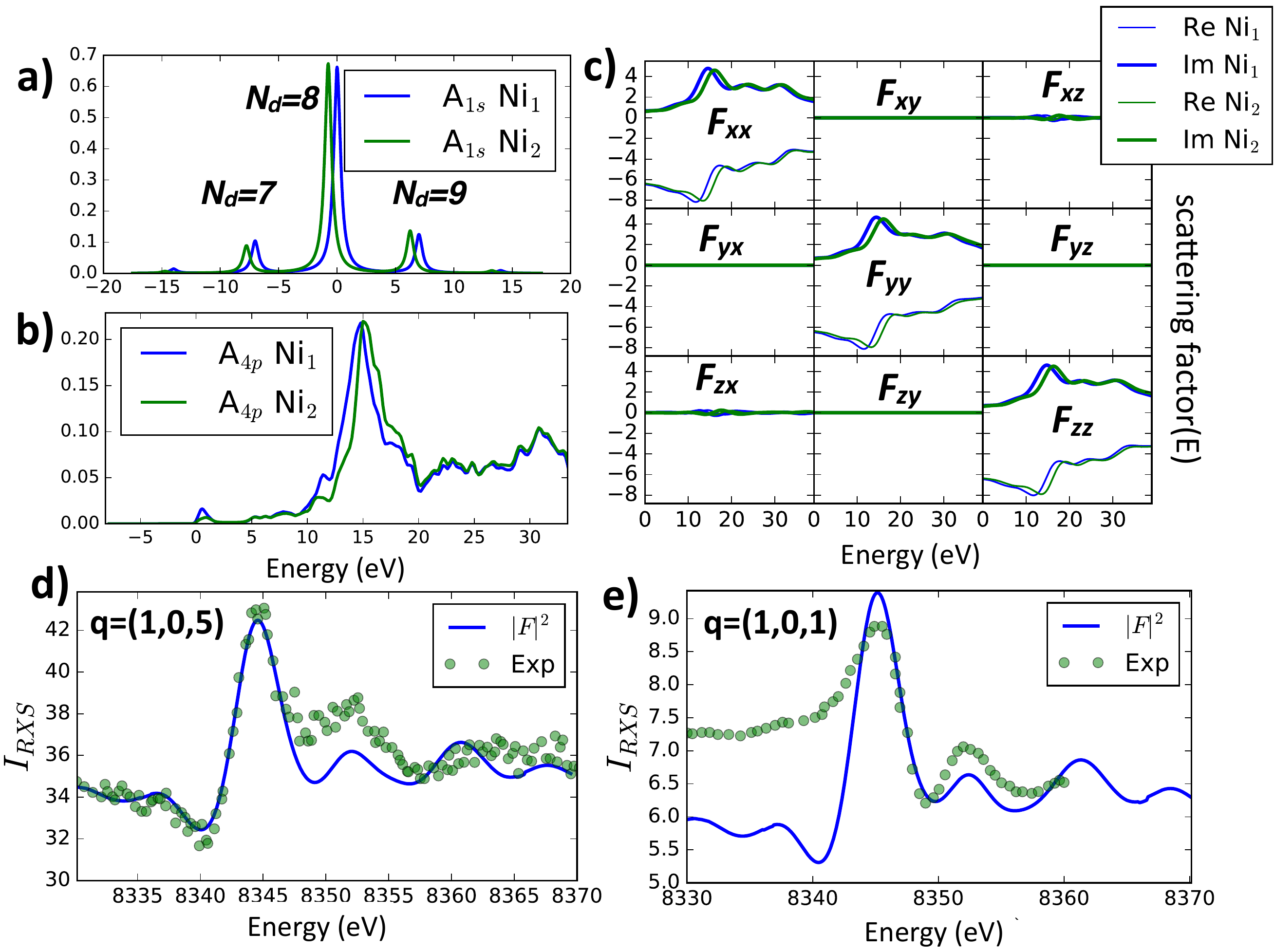}
\caption{ \textbf{Resonant Elastic X-ray scattering on Ni K-edge:}  
a) The spectral function of the $1s$ core state in the presence of
the fluctuating valence of the Ni $3d$ shell. b) Ni $4p$ density of
state. c) the energy dependent matrix of the scattering factor, where
E means electron units d-e)
measured and computed X-ray scattering intensity at the two Bragg
peaks.  Experimental data in d) are reproduced from
Ref.~\onlinecite{Xray1} and in e)
from Ref.~\onlinecite{Keimer_RXS}.
}
\label{fig3}
\end{figure}
The CD model was originally invented to explain
the RXS results~\cite{Xray1},
which showed a strong energy dependent signal at the weak nuclear 
Bragg peak $(h,k,l)$, where $h+k+l$ is even, and $l$ is odd. 
If the scattering factor $f$
of each Ni atoms is approximated by a spherically symmetric quantity,
the resonant part of the structure
factor is  directly proportional to the difference
$f_{Ni_1}-f_{Ni_2}$~\cite{suppl}
A strong X-ray
signal at the Ni resonance can therefore be taken as a direct evidence
of a very large difference between the two Ni atoms. In particular,
for the X-ray K-edge measurement, this must mean that the energy
difference between the core $1s$ state and the valence $4p$ states of
the Ni ion is very different on the two inequivalent Ni sites. As the
core energy is very sensitive to the amount of the charge on Ni ion,
it is generally accepted that  the difference in the core energy
comes from the different charge accumulated on Ni$_1$ and Ni$_2$
ions. As our model predicts negligible total charge difference on the two
Ni sites, the X-ray scattering needs an alternative explanation.

In Fig.~\ref{fig3}a and b we show the calculated spectral function for
the Ni $1s$ core and $4p$ valence orbital. The $1s$ spectra on Ni$_1$
ion is shifted up compared to Ni$_2$ for approximately $0.7\,$eV, and
the $4p$ spectra is shifted in opposite direction for approximately
$0.8\,$eV, resulting in approximately 1.5$\,$eV difference in the
$1s\rightarrow 4p$ transition energy on two inequivalent Ni atoms.
Such energy difference can explain the occurence of the main peak in
the RXS intensity displayed in
Fig.~\ref{fig3}d-e, hence no charge order is needed for its
explanation. However, the multiple peak structure of the intensity can
not be explained by only the single-particle effects and the
structural distortion.
The $4p$ states are very extended and do not appreciably overlap with
the core $1s$ states, hence the Coulomb repulsion between the two can
be neglected. However, the Ni $1s$ and partially filled $3d$ orbitals 
have large overlap, therefore the Coulomb repulsion between the two
states is comparable to the Coulomb $U$ among electrons in the $3d$
shell. In this work, we included such core-hole interaction between
the Ni $1s$ and Ni $3d$ states, which takes the form
$\Delta H = U_{ch} (n_{1s}-2)(n_{3d}-\braket{n_{3d}})$.  We took
$U_{ch}=7\,$eV, the same as $U$ in the $3d$ shell. When such term is
included in the Hamiltonian, the core $1s$ orbital experiences
different energy when the $3d$ shell is in different valence state. As
there are substantial valence fluctuations in this system with finite
probability for Ni $d^7$ and $d^9$ valence, the core state spectra is
split into three peaks, roughly separated by $U_{ch}$.  Finally, the
scattering factor on each Ni, is computed by convoluting $1s$ and $4p$
spectra (see supplementary chapter IV) and is displayed in
Fig.~\ref{fig3}c. The $xy$ and $yz$ components vanish by the symmetry,
and only diagonal and the $xz$ components are finite. Moreover, the
$xz$ component is one order of magnitude smaller than the diagonal
components, as consistent with the fact that the $\sigma-\pi$
intensity is two orders of magnitude smaller than $\sigma-\sigma$
scattering
intensity~\cite{spherical_symmetric,spherical_symmetric_equation} (
the off-diagonal component would contribute to the scattering in
$\sigma-\pi$ channel).
Moreover,
the diagonal components have a pre-peak shoulder roughly $U_{ch}$
below the main peak, and second and third peak roughly $U_{ch}$ and
$2U_{ch}$ above the main peak, all consequence of the core-hole
interaction. Finally, computing the square of the total structure factor
we arrive at the X-ray
intensity, displayed in Fig.~\ref{fig3}d and e. This is directly
compared with the experiment, and we notice that reasonable agreement
is achieved without any fitting parameter.  We can thus conclude that
the inequivalent Ni-sites harboring "site-selective Mott transition''
but no real charge order, can explain all important observation in the
rare-earth nickelates, including the magnetic long range order
consistent with neutron scattering data, and the resonant X-ray
intensity in the weak nuclear Bragg peaks, which was previously assumed
to be a proof of the electronic charge order.

\section*{Acknowledgements}
G.L.P. was supported by the  U.S. Department of Energy, Office of Science, 
Basic Energy Sciences, as a part of the Computational Materials Science Program, funded by the  U.S. Department of Energy,
Office of Science, Basic Energy Sciences, Materials Sciences and
Engineering Division.
K.H. work on RXS was supported by NSF DMR-1405303 and the work on
forces and free energy was supported by
U.S. Department of Energy as a part of the above mentioned Computational Materials
Science Program.

\section*{Author contributions}
K.H and L.P. carried out the calculations. K.H. developed the DMFT
code. K.H. and L.P. analysed the results and wrote the paper. 

\section*{Competing financial interests}
The authors declare no competing financial interests.

\bibliographystyle{naturemag}
\bibliography{refNi}

\end{document}